\documentclass[twocolumn,showpacs,superscriptaddress,prl,aps,floatfix]{revtex4}

\usepackage{graphicx}
\usepackage{rotating}
\usepackage{amsmath}

\newcommand{\be}{\begin{equation}}
\newcommand{\ee}{\end{equation}}
\newcommand{\bea}{\begin{eqnarray}}
\newcommand{\eea}{\end{eqnarray}}
\newcommand{\ba}{\begin{array}}
\newcommand{\ea}{\end{array}}

\newcommand{\aF}{\alpha^2 F(\Omega)}
\newcommand{\bk}{{\bf k}}
\newcommand{\bkp}{{\bf k'}}
\newcommand{\EF}{$E_F$}
\newcommand{\D}{$\Delta$}
\newcommand{\Ds}{$\Delta_\sigma$}
\newcommand{\Dp}{$\Delta_\pi$}
\newcommand{\nk}{{n\bf k}}
\newcommand{\nkp}{{n'\bf k'}}

\newcommand{\s}{$\sigma$}
\newcommand{\p}{$\pi$}
\newcommand{\mgbtwo}{MgB$_2$}
\newcommand{\tc}{$T_{\rm c}$}

\begin{document}

\title{Superconducting properties of MgB$_2$ from first principles}

\author{A.~Floris}
\affiliation{Institut f{\"u}r Theoretische Physik, Freie Universit{\"a}t Berlin, Arnimallee 14, D-14195 Berlin, Germany}
\affiliation{SLACS (INFM), Sardinian Laboratory for Computational Materials Science}
\affiliation{Dipartimento di Scienze Fisiche, Universit\`a degli Studi di Cagliari,
S.P. Monserrato-Sestu km 0.700, I--09124 Monserrato (Cagliari), Italy}

\author{G.~Profeta}
\affiliation{C.A.S.T.I. (INFM), Center for Scientific and Technological Assistance to
Industries}
\affiliation{Dipartimento di Fisica, Universit\`a degli studi dell'Aquila,
I-67010 Coppito (L'Aquila) Italy}

\author{N.\,N.~Lathiotakis}
\affiliation{Institut f{\"u}r Theoretische Physik, Freie Universit{\"a}t Berlin, Arnimallee 14, D-14195 Berlin, Germany}

\author{M.~L{\"u}ders}
\affiliation{Daresbury Laboratory, Warrington WA4 4AD, United Kingdom }

\author{M.\,A.\,L.~Marques}
\affiliation{Institut f{\"u}r Theoretische Physik, Freie Universit{\"a}t Berlin, Arnimallee 14, D-14195 Berlin, Germany}

\author{C.~Franchini}
\affiliation{SLACS (INFM), Sardinian Laboratory for Computational Materials Science}
\affiliation{Dipartimento di Scienze Fisiche, Universit\`a degli Studi di Cagliari,
S.P. Monserrato-Sestu km 0.700, I--09124 Monserrato (Cagliari), Italy}

\author{E.\,K.\,U.~Gross}
\affiliation{Institut f{\"u}r Theoretische Physik, Freie Universit{\"a}t Berlin, Arnimallee 14, D-14195 Berlin, Germany}

\author{A.~Continenza}
\affiliation{C.A.S.T.I. (INFM), Center for Scientific and Technological Assistance to
Industries}
\affiliation{Dipartimento di Fisica, Universit\`a degli studi dell'Aquila,
I-67010 Coppito (L'Aquila) Italy}

\author{S.~Massidda}
\altaffiliation{Also at LAMIA-INFM, Genova, Italy}
\affiliation{SLACS (INFM), Sardinian Laboratory for Computational Materials Science}
\affiliation{Dipartimento di Scienze Fisiche, Universit\`a degli Studi di Cagliari,
S.P. Monserrato-Sestu km 0.700, I--09124 Monserrato (Cagliari), Italy}

\begin{abstract}
Solid \mgbtwo\   has rather interesting 
and technologically important properties, such as a very high superconducting transition 
temperature. Focusing on this compound, we report the first non-trivial application of a
novel density-functional-type theory for superconductors, recently proposed by the authors.
Without invoking any adjustable parameters, we obtain the transition temperature, the gaps, 
and the specific heat of \mgbtwo\ in very 
good agreement with experiment. Moreover, our calculations show how the Coulomb interaction 
acts differently on \s\ and \p\ states, thereby stabilizing the observed superconducting
phase.
\end{abstract}
\maketitle
Understanding and predicting superconducting properties of real materials is of both
fundamental and technological importance. While there are interesting classes of
materials, such as the high \tc\ Cu oxides, where the superconducting mechanism is still
under debate, the recent discovery of phonon-mediated superconductivity at 39.5\,K in 
\mgbtwo\cite{akimizu} -- but also in other materials -- 
keeps the interest in the phonon-driven mechanism alive. \mgbtwo\ has
rather peculiar properties, such as the presence of two 
superconducting gaps at the Fermi level. While two-band or, more generally, multi-band superconductivity 
has long been known \cite{suhl59} to favor a high critical temperature, there remains 
the challenging question: Could
anyone have predicted quantitatively the peculiar superconducting phase in \mgbtwo, 
including its high \tc value, by means of ``physically unbiased'', first-principles
calculations? 

As a matter of fact, despite the theoretical and technological interest involved, the
capability to predict from first principles material-specific properties,
such as the critical temperature and the gap, has been out of reach so far. This is 
because (conventional) superconductivity appears as the 
result of a subtle competition between two opposite effects, the phonon mediated 
attraction (denoted ``e-ph'' in the following) and the direct Coulomb repulsion (denoted ``e-e'') 
between the electrons. 
Historically, after the microscopic identification of the superconducting order parameter
by Bardeen, Cooper and Schrieffer (BCS) \cite{bcs}, the first theoretical framework 
aiming at the description of real materials was put forth by Eliashberg 
\cite{eli}. In this theory the e-ph 
interaction is perfectly accounted for, however the effects of the e-e Coulomb
repulsion are condensed in a single parameter, $\mu^*$, which is difficult to 
calculate from first principles and which, in most practical applications, is treated
as an adjustable parameter, usually chosen as to reproduce the experimental \tc. In
this sense, Eliashberg theory, in spite of its tremendous success, has to be
considered a semi-phenomenological theory. 

Looking at normal-state properties, density functional theory (DFT) \cite{dreizler90} has enjoyed
increasing popularity as a reliable and relatively inexpensive tool to 
describe real materials. In 1988 the basic concepts of DFT were generalized
to the superconducting phase by Oliveira, Gross and Kohn (OGK) \cite{ogk} via
including the superconducting order parameter as an additional "density" in
the formalism. The theory of OGK, however, was restricted to weak e-ph
coupling. Along the lines of a recently presented multi-component DFT \cite{kreibich}
of electrons and ions, the theory of OGK was successfully generalized to the
strong-coupling case \cite{PRB}. In this Letter we apply the strong-coupling DFT 
to the challenging case of \mgbtwo.

The central equation of the DFT for superconductors is a generalized gap equation 
of the form 
\begin{equation} 
  \label{eq:gap} 
  \Delta_\nk = - {\cal Z}_\nk \;\Delta_\nk 
  -\frac{1}{2}\sum_\nkp {\cal K}_{\nk,\nkp}
  \frac{\tanh\left(\frac{\beta}{2}E_\nkp\right)}{E_\nkp} \; \Delta_\nkp,
\end{equation}
where $n$ and $\bk$ are respectively the electronic band index and the wave vector 
inside the Brillouin zone. $\beta$ is the inverse temperature and 
$E_\nk=\sqrt{(\varepsilon_\nk-\mu)^2+ \left|\Delta_\nk\right|^2}$ 
are the excitation energies  of the superconductor, defined in terms of the gap 
function $\Delta_{\nk}$, the Kohn-Sham eigenenergies of the normal state 
$\varepsilon_\nk$, and the chemical potential $\mu$. The kernel, ${\cal K}$, 
appearing in the gap equation consists of two contributions 
${\cal K}={\cal K}^{\rm e-ph}+{\cal K}^{\rm e-e}$, representing the effects of 
the e-ph and of the e-e interactions, respectively. The first of these terms
involves the e-ph coupling matrix, while the second contains the matrix elements 
of the screened Coulomb interaction. Eq.~(\ref{eq:gap}) has the same structure 
as the BCS gap equation, with the kernel ${\cal K}$ replacing the model interaction 
of BCS theory. This similarity allows us to interpret the kernel as an effective 
interaction responsible for the binding of the Cooper pairs. On the other hand, 
${\cal Z}$ plays a similar role as the renormalization term in the Eliashberg 
equations. We emphasize that Eq.~(\ref{eq:gap}) is not a mean-field equation (like 
in BCS theory), since it contains correlation effects. Furthermore, it has the 
form of a static equation -- i.e., it does not depend explicitly on the frequency --  
and therefore has a simpler structure than the Eliashberg equations. However, this 
certainly does not imply that retardation effects are absent from the theory: as a 
matter of fact, an Eliashberg-type spectral function $\aF$ enters the calculation 
of ${\cal Z}$ and ${\cal K}^{\rm e-ph}$.

Coming back to \mgbtwo,  its Fermi surface  has several sheets with different orbital 
character (see e.g. Ref.~\onlinecite{kortus}). In particular, the tubular 
structures with \s\ character are very strongly coupled to the $E_{2g}$ phonon mode, 
corresponding to a B-B bond-stretching in the boron planes \cite{kortus,PickettKong}.
\mgbtwo\  also has three-dimensional 
\p\ bands, that give rise to a complicated Fermi surface. Without holes in the \s\  bands, the compound 
would not be superconducting, like  AlB$_2$. The \p\  bands 
are coupled much less efficiently to phonons, but are nevertheless crucial to 
superconductivity. A remarkable feature of this compound is the presence of two gaps on 
the \s\  and \p\  bands, as clearly demonstrated by several different experiments 
\cite{iavarone,szabo,schmidt, gonnelli,TUNNEL,putti,bouquet,yang}. 
On the theoretical side, this system has been treated 
within the {\bf k}-resolved Eliashberg theory \cite{choi,choi_nat}, using a two-band 
scheme \cite{Liu}, with four e-ph spectral functions to represent the distinct 
couplings. 
Correspondingly, the anisotropy of the Coulomb interaction has also been 
investigated\cite{mazin2004,moon}, with  $\mu^*$  treated as a matrix. 

In our density functional calculations we used the four, band resolved, Eliashberg 
functions, $(\alpha^{2}F_{n,n'}(\Omega) ; n,n'=\sigma,\pi)$,
previously employed within a two-band Eliashberg scheme by Golubov {\it et al.} \cite{al2f_mgb2}. 
Our procedure keeps the fundamental distinction between \s\ and \p\ gaps,
analogously to the Eliashberg calculations reported to date. We recall here that a fully 
consistent calculation should not use the $\alpha^{2}F_{n,n'}(\Omega)$ functions,
but rather the $\nk,\nkp$-resolved couplings:  By using the $\alpha^{2}F_{n,n'}(\Omega)$, 
the e-ph interaction is averaged over {\bf k} and {\bf k'} at the FS, which may have a 
small, but non-negligible effect. We define the bands crossing the FS to be of \s\ 
character if they are contained in a cylinder of basis radius $1/4$ of a reciprocal 
lattice vector, and of \p\ character otherwise. Away from the Fermi surface this distinction is 
meaningless but harmless, as the phonon terms die off quickly and the Coulomb term is 
independent of this distinction. As we described our computational strategy elsewhere 
\cite{PRB}, we shall not report the details here. We just mention that extreme care 
needs to be taken with the sampling of the region in $\bk$-space around the FS. We 
accomplish this task by using around 65000 independent $\bk$ points, chosen according to 
a Metropolis algorithm. This places $\approx 8000$ points within $\approx \hbar \omega_{E_{2g}}$ 
of the FS. We estimate the overall numerical stability of our results to be roughly 5\%.

\begin{figure}
\begin{center}
\includegraphics[width=0.4\textwidth,clip]{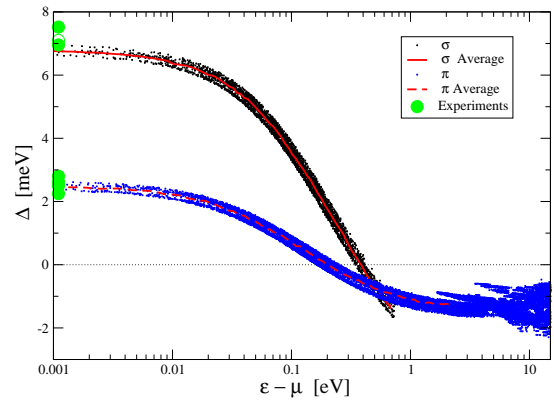}
\end{center}
\caption[]{Calculated superconducting gap of \mgbtwo\ as a function of energy ($T=0$\,K). }
\label{gapE}
\end{figure}

As for the e-e interaction, we calculated the matrix elements of the screened Coulomb 
potential with respect to the Bloch functions of \mgbtwo, for the whole energy range 
of relevant valence and conduction states, using the full-potential linearized 
augmented plane wave method. In our previous work \cite{PRB}, the Coulomb 
interaction was screened using a simple Thomas-Fermi model. However, the different 
nature of \s\ and \p\ bands in \mgbtwo, and in particular the highly localized 
character of the former, strongly calls for the use of a non-diagonal screening, 
including local field effects. To avoid the cumbersome calculation of the 
dielectric matrix of \mgbtwo, we used a dielectric matrix obtained with the method 
of Hybertsen and Louie \cite{HL_DM}. In particular, we substituted the 
model dielectric function in Eq.~(7) of Ref.~\onlinecite{HL_DM} by a Thomas-Fermi 
model, computed at the local (valence) density of \mgbtwo. The diagonal part of our 
model compares rather well with the calculations of Zhukov {\em et al.} \cite{QTFref}.  

In Fig.~\ref{gapE} we plot the energy gap as a function of (positive) energy distance from the 
Fermi energy (the gap function 
exhibits particle-hole symmetry to a good extent). We can see that the two gaps of 
the  material, \Ds\ and \Dp, appear naturally from our calculations. The \s\ gap 
is defined only up to the  energy of the top of the \s\ band. Both \Ds\ and \Dp\ 
are anisotropic. This results from the anisotropy of the Coulomb potential matrix 
elements -- roughly 0.4\,meV, $\approx$ 6\% of \Ds\ at the FS and gets much larger at high energy, 
where there are many bands with different 
characters. The averages of \Ds\ and \Dp\ at the Fermi level (6.8\,meV and 2.45\,meV 
respectively) are in excellent agreement with experiment. Both gaps 
change sign, which is a necessary condition to find superconductivity in the presence 
of the repulsive Coulomb interaction. In fact, our gap equation does not converge to a 
superconducting solution, unless all electronic states in a large energy window are 
included.

\begin{figure}
\begin{center}
\includegraphics[width=0.4\textwidth,clip]{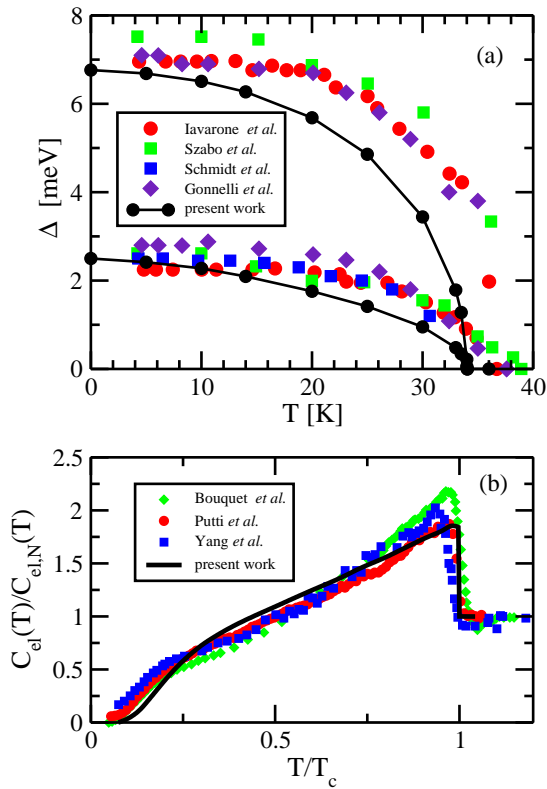}
\end{center}
\caption[]{Superconducting gaps at the FS and specific heat of \mgbtwo\ .
Panel ({\bf a}): Comparison between theoretical and experimental gap 
at the FS plotted as a function of temperature. The calculated gaps and
 $T_{\rm c}$ (34.1\,K) are obtained without the use of any adjustable
parameter. Panel ({\bf b}): Experimental and calculated electronic specific heat, 
as a function of  $T/T_{\rm c}$. }
\label{gapT}
\end{figure}

In Fig.~\ref{gapT} (panel ({\bf a})), the superconducting gaps are plotted versus temperature, 
together with a few recent 
experimental results. The agreement is striking: the values of \tc\ (34.1\,K)  and of 
\Ds\  and \Dp\  at $T=0$\,K are very close to the experimental data. Moreover, the 
temperature behavior of both gaps, along with their strongly non-BCS behavior, are very 
well reproduced. We believe that such an agreement for a highly non-trivial superconductor 
such as \mgbtwo, without any adjustable parameter, is unprecedented in the field of 
superconductivity. Obviously, unlike calculations performed using Eliashberg theory, we 
do not reproduce exactly the experimental critical temperature, as our calculations are 
not fitted to match any experimental quantity. 

We also calculated the Kohn-Sham entropy as a function of temperature and, from its 
temperature derivative, the specific heat. In order to  compare our results with 
experiments \cite{putti,bouquet,yang}, we plot in Fig.~\ref{gapT} (panel ({\bf b})) 
the reduced specific heat versus temperature, normalized to \tc\ (using the corresponding 
experimental and calculated \tc\ values). Both the shape of 
the curve as well as the discontinuity at \tc\ are almost perfectly reproduced. 
We recall that the   low temperature shoulder comes 
from the presence of the smaller \p\ gap and that our \Ds$/$\Dp\ is slightly different from 
the experimental ratio.  

While the good agreement with experiment  underlines the predictive
power of our method, it is only one part of our investigation. Another 
important aspect is to gain further insight into the peculiar superconductivity of \mgbtwo.
To this end, we performed a calculation using an average functional form 
for the Coulomb interaction, not distinguishing between \s\ and \p\ bands.  This 
functional, described in detail in Ref.~\onlinecite{PRB}, corresponds to a 
semiclassical treatment valid in the limit of slowly varying densities, and actually 
leads to a good agreement with the full matrix element calculation for $s,p$-metals, 
and works reasonably also in $d$-metals as Nb. We obtain $T_c=$52~K, with the \s\ and 
\p\ gaps being 9.8~meV  and 1.9~meV respectively. This test shows  that the repulsion 
among \s\ states, stronger than within \p \ and between \s\ and \p, is crucial in 
achieving good agreement with experiment. We also see clearly that the more delocalized 
\p\ electrons are reasonably well described by an average formula derived from 
free-electron concepts, while the repulsion among \s\ electrons needs the real matrix 
element calculations.  Reversely, if we average all the e-ph spectral functions, we 
obtain a single gap \D=3.7~meV and the much lower $T_c\approx 20 $~K,
in agreement with the analysis of Ref. \cite{Liu} and
with a similar test carried out by Choi {\it et al.}\cite{choi}. 

\begin{figure}
\begin{center}
\includegraphics[width=0.4\textwidth,clip]{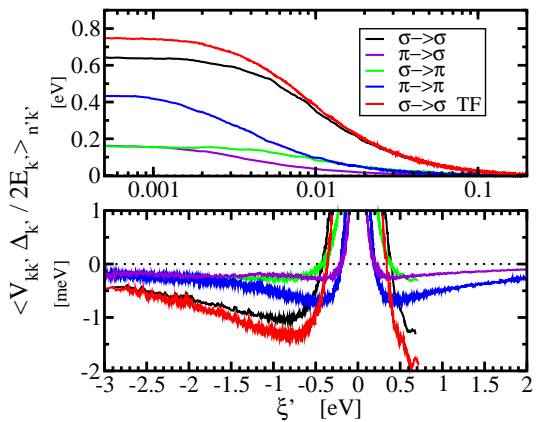}
\end{center}
\caption[]{Electronic repulsion contributions to the gap equation (see 
Eq.~\ref{eq:gap}), at an arbitrary \bk, averaged over \bkp\  and bands on 
isoenergy surfaces, but keeping the \s, \p\  distinction.}
\label{MEL}
\end{figure}

Of course, in the search for novel superconductors with higher transition temperatures, it 
would be desirable to keep the extremely strong e-ph coupling for \s\ states, while reducing 
the corresponding Coulomb interaction to that of more delocalized \p\ states. Unfortunately,  
the two features are linked together, as both the strong e-ph coupling and the strong 
Coulomb repulsion derive from the covalent nature of the corresponding electronic states.

To push this analysis further, we  plot, in the upper panel of Fig.~\ref{MEL}, the energy  
dependence of the Coulomb contribution to Eq.(\ref{eq:gap}) at $T\approx 0 \rm K$, namely
${\cal K}^{el}_{\nk,\nkp} \Delta_\nkp/(2E_\nkp)\equiv {\cal K}^{el}_{\nk,\nkp} \chi_\nkp$ 
against $\xi_\nkp=\varepsilon_\nkp-E_F$, for a few $\bk$-points arbitrarily chosen on the 
Fermi surface. As $\chi_\nkp$ goes to 0.5 on the 
Fermi surface, Fig.~\ref{MEL} shows the larger magnitude of the intraband \s-\s\ and \p-\p, 
relative to the interband \s-\p\ matrix elements (see also Ref. \onlinecite{moon}). 
The different behaviour of the \s $\to$ \p\  and \p $\to$ \s\  
terms results from the $\chi$ factor, as the matrix elements themselves are symmetric. Obviously, 
the \s-\s\ repulsive matrix elements are the strongest. The scattering of data for a given 
energy comes from the different orbital character of wavefunctions at $\nkp$.  The energy 
dependence of the quantities plotted in Fig.~\ref{MEL} comes almost entirely from $\chi_\nkp$, 
as the matrix elements themselves have a marginal energy dependence. 
In order to show how the reduction of Coulomb repulsion takes place, we emphasize in the lower 
panel of Fig.~\ref{MEL} the region corresponding to low matrix elements (on a linear scale).
We  see how,   when  \D\ becomes negative, the Coulomb interaction actually gives a constructive
contribution due to the minus sign in Eq.~(\ref{eq:gap}). Although the negative values are 
much smaller (by almost 3 orders of magnitude) than the positive ones, the corresponding energy 
range  is much larger, resulting into the well known, substantial reduction of the effective 
Coulomb contribution.  The most important energy region is located 
in between 0.50 to 3 eV below \EF, in particular due to the 
strong intraband \s-\s\  matrix elements. The interband contribution from \p\ bands (violet in 
Fig.~\ref{MEL}), on the other hand, is considerably smaller, which is obviously the case also 
for the \s\ contribution to \Dp\  (green in Fig.~\ref{MEL}). Summing up over $\nkp$, the 
negative contribution to \Ds\ coming from the (positive gap) region of the \s\ Fermi surface 
overcomes by a factor of $\approx 7$ the contribution coming from the corresponding \p\ region.

Finally, it is also interesting to investigate the importance of local field (LF) effects on 
the superconducting properties of \mgbtwo. It turns out that using a diagonal Thomas-Fermi (TF) 
screening (which completely neglects LF's) the \s-\s\ matrix elements increase by roughly 
15\% (red in Fig.~\ref{MEL}), while the \s-\p\ and \p-\p\ terms remain basically unchanged. As a consequence, neglecting 
LF effects leads to a marginal ($\approx 4$\%) decrease of \Dp, but decreases significantly 
\Ds\ (by about 14\%). The different  behavior of \Ds\ and \Dp\ can be understood quite 
naturally: LF corrections imply a better screened interaction among electrons when they are 
located in a high density region inside the unit cell. This is precisely the case of the \s\ 
bands. On the other hand, the \p\ bands are more delocalized -- the electrons reside in the 
interstitial region -- and are therefore reasonably described by diagonal screening.

In this communication we presented the first non-trivial application of a recently developed 
ab-initio theory of superconductivity. In particular, we obtained for \mgbtwo\  the value of \tc, 
the two gaps, as well as the specific heat as a function of temperature in very good agreement 
with experiment. We stress the predictive power of the approach presented: being, by its very 
nature, a fully ab-initio approach, it does not require semi-phenomenological parameters, such 
as $\mu^*$. Nevertheless, it is able to reproduce with good accuracy superconducting properties, 
up to now out of reach of first-principles calculations. Furthermore, our calculations allow for 
a detailed analysis of the contribution of the Coulomb repulsion to the  superconducting gap, 
opening the way to tailoring the electronic properties of real materials in order to optimize 
superconducting features.

\begin{acknowledgments}
Work supported by the italian INFM (PRA UMBRA 
and  a  grant at Cineca, 
Bologna, Italy), 
 by the Deutsche Forschungsgemeischaft within the program SPP 1145, by the EXC!TiNG
Research and Training Network of the European Union, and by the NANOQUANTA NOE.
S.M. aknowledges A. Baldereschi and G. Ummarino for stimulating discussion. 
\end{acknowledgments}

\end{document}